\documentclass[review]{elsarticle}

\usepackage{hyperref}
\usepackage[usenames]{color}

\begin{document}

\begin{frontmatter}

\title{Suppression of long range magnetic ordering and electrical conduction in $Y_{1.7}Bi_{0.3}Ir_2O_7$ thin film}


\author{Vinod Kumar Dwivedi \textit* and Soumik Mukhopadhyay}
\address{\textit{$^{1}$}Materials Science Programme, Indian Institute of Technology Kanpur, Kanpur 208016, India}
\address{\textit{$^{2}$}Department of Physics, Indian Institute of Technology Kanpur, Kanpur 208016, India}


\cortext[mycorrespondingauthor]{Corresponding author}
\ead{vinodd@iitk.ac.in}

\begin{abstract}
We find that the long range magnetic ordering is absent and electrical conduction suppressed in $Y_{1.7}Bi_{0.3}Ir_2O_7$/YSZ(100) thin film prepared by pulsed laser deposition. The sharp down-turn of inverse magnetic susceptibility $\chi^{-1}(T)$ from the conventional Curie-Weiss behaviour below $T^\ast \sim 168 K$ suggests an inhomogeneous ferromagnetic Griffiths like phase. The transport and magnetic properties are explained on the basis of the coexistence of mixed oxidation states of Ir, (i.e. $Ir^{4+}$ and $Ir^{3+}$) leading to non-magnetic defects and reduction in $t_{2g}$ density of states at the Fermi level. 
\end{abstract}


\end{frontmatter}


\section{Introduction}
The interplay between comparable energy scales, namely the strong relativistic spin-orbit coupling (SOC), crystal field effect and on site Coulomb repulsion giving rise to novel topological phases makes the 5d iridate systems fundamentally interesting~\cite{Pesin,Kim,Machida,Wan,Krempa,Abhishek1,Abhishek2}, with promising applications in quantum computing and spin based electronics~\cite{Hasan}. The ratio of the energy scales between electronic correlation and kinetic hopping can be effectively tuned by changing the A-site ionic radius in pyrochlore iridates with general formula $A_2Ir_2O_7$ (where A = Y, Bi or rare earth ions)~\cite{Krempa}, leading to metal insulator transition which is often accompanied by a paramagnetic to an antiferromagnetic (AFM) phase transition~\cite{Yanagishima}.

While several interesting results have been reported in bulk A-227 compounds,~\cite{Yanagishima,Aito,Soda,Liu,Taira,Harish,Zhu,Vinod,Vinod1,Vinod2,Vinod3} experimental reports on thin films of pyrochlore iridates are few and far between~\cite{Fujita,Fujita1,Fujita2,Gallagher,Bing,Ohtsuki,Yang1,
Yang3,Woo}. Theoretical studies on pyrocholore iridate thin films have suggested a variety of emergent topological properties, such as quantized anomalous Hall conductance~\cite{Bohm}, unconventional time-reversal symmetry broken superconducting state~\cite{Laurell} and antiferromagnet (AF) semimetals with and without Weyl points~\cite{Ishii}. Thin films of iridates can provide an ideal template to study these emergent properties arising from lattice mismatch induced strain and finite size effects. In this manuscript, we discuss the structural, magnetic and electrical transport properties of $Y_{1.7}Bi_{0.3}Ir_2O_7$ thin films deposited on YSZ(100) substrates.

\section{Experimental methods}
The bulk polycrystalline (S2) sample of $Y_{1.7}Bi_{0.3}Ir_2O_7$ was prepared via a solid state reaction route following the protocol reported elsewhere~\cite{Aito}. This sample will be, henceforth, used as a reference. Thin film (S1) of $Y_{1.7}Bi_{0.3}Ir_2O_7$ was deposited on YSZ(100) substrate using a $KrF$ pulsed excimer laser with wavelength $\lambda$ = 248 nm. The repetition rate, laser fluence, substrate temperature and oxygen partial pressure were kept at 5Hz, 3.3J/cm$^2$, 500$^o$C and 100 mTorr, respectively. After deposition, the PLD chamber was filled with oxygen gas
up to atmosphere pressure, in which thin film was
cooled down. Post-deposition annealing was done at 1000$^o$C for 2h in air. X-ray diffraction (XRD) patterns were measured using a Philips XPert-Pro MPD X-ray diffractometer with CuK$\alpha$ radiation ($\lambda$= 1.54$\AA$) at room temperature. Energy dispersive x-ray spectroscopy (EDS) was carried out using a JSM-7100F, JEOL field emission scanning electron microscope (FE-SEM) to determine the chemical composition and compositional uniformity. An atomic force microscope (Asylum Research AFM, MFP3D) was used for topographical characterization. X-ray photoelectron spectroscopy (XPS) data were taken on a PHI 5000 Versa Probe II system. The magnetic measurements were done by Quantum Design Physical Property Measurement System (PPMS).

\section{Results and Discussion}
\begin{figure}
\includegraphics[width=9cm]{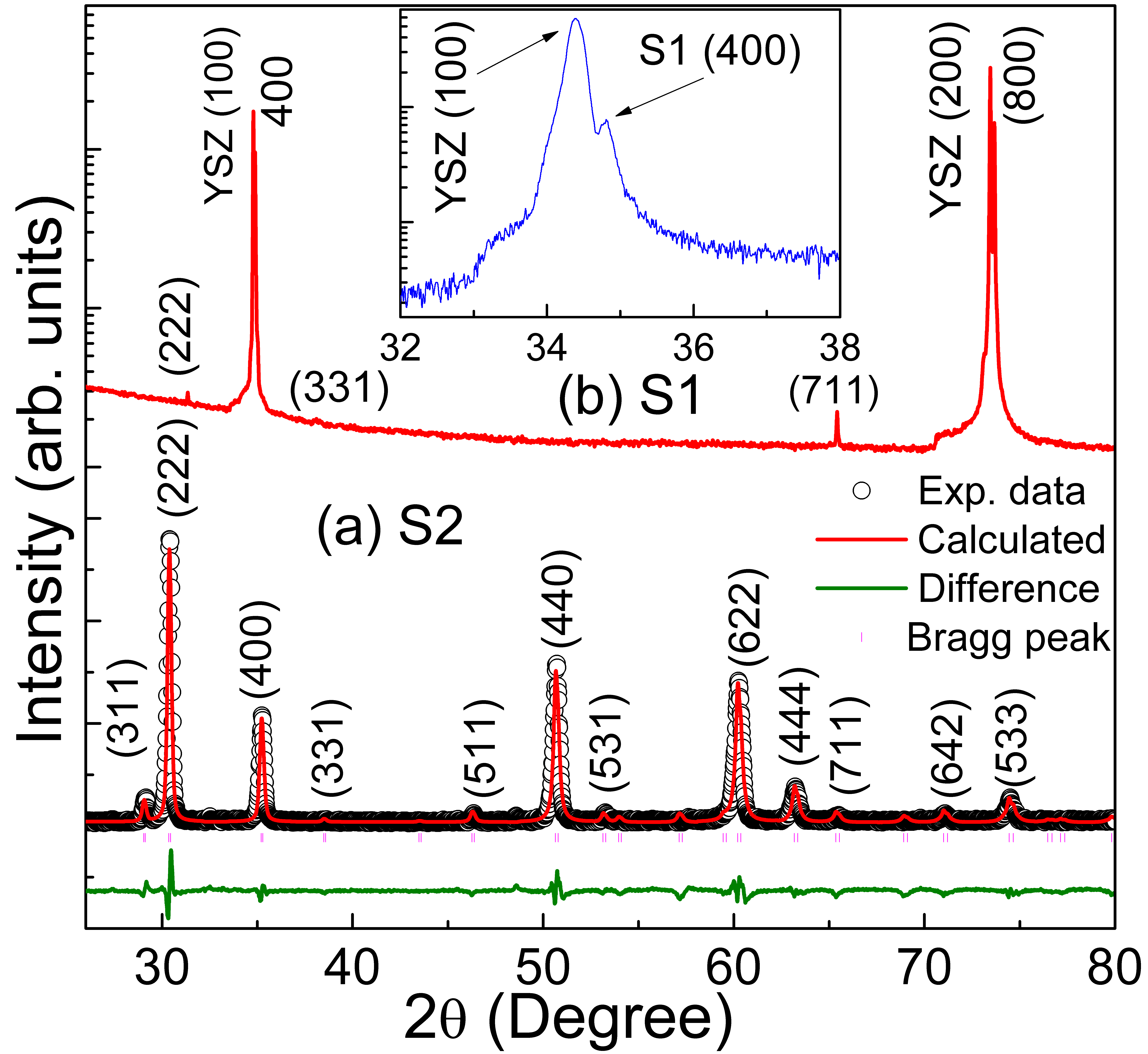}\\
\caption{XRD spectra of (a) bulk polycrystalline sample S2 and (b) thin film S1; inset shows high resolution XRD scan of the S1 around the YSZ (100) and S1(400) peaks.}\label{fig:xrd}
\end{figure}

Fig.~\ref{fig:xrd}a shows the powder XRD data of S2 analyzed using Rietveld method with Fullprof refinement program which suggests oriented crystalline pyrochlore cubic phase with space group Fd-3m except small fraction of impurity phases. The obtained lattice constant for S2 is $a = 10.20 \AA$. A $\theta-2\theta$ XRD scan of thin film S1 is shown in Fig.~\ref{fig:xrd}b. A well-defined thin film (400) peak is observed (inset, Fig.~\ref{fig:xrd}b), with a shift towards the higher angle [$\bigtriangleup 2\theta \sim 0.7^0$] as compared to S2. The appearance of peaks such as (331), (711) etc suggests strain-relaxed polycrystalline nature of film. The grain size can be estimated using broadening in either of the two above [(222), (711)] mentioned peaks. The average grain size turns out to be 300 nm. The in-plane d-spacing corresponding to the (400) Bragg peak turns out to be $d_{400}\sim 2.59 \AA$, which is 1.6$\%$ larger than that for the bulk sample S2 ($\sim 2.55 \AA$). The lattice constant for YSZ (100) orientation is $2a_{YSZ} = 10.25 \AA$ with $d_{100}$ being $\sim 2.60 \AA$. The enhancement of lattice parameter in S1 is likely due to strain from the YSZ substrate. We have calculated the strain in the thin film using the relation $S = \frac{d_f-d_s}{d_s} \times 100$, where $d_f$ and $d_s$ are the d-spacing for the film and substrate, respectively. The obtained value of $S$ is -0.4$\%$, which suggests biaxial tensile strain present in the film.

\begin{figure}
\includegraphics[width=8cm]{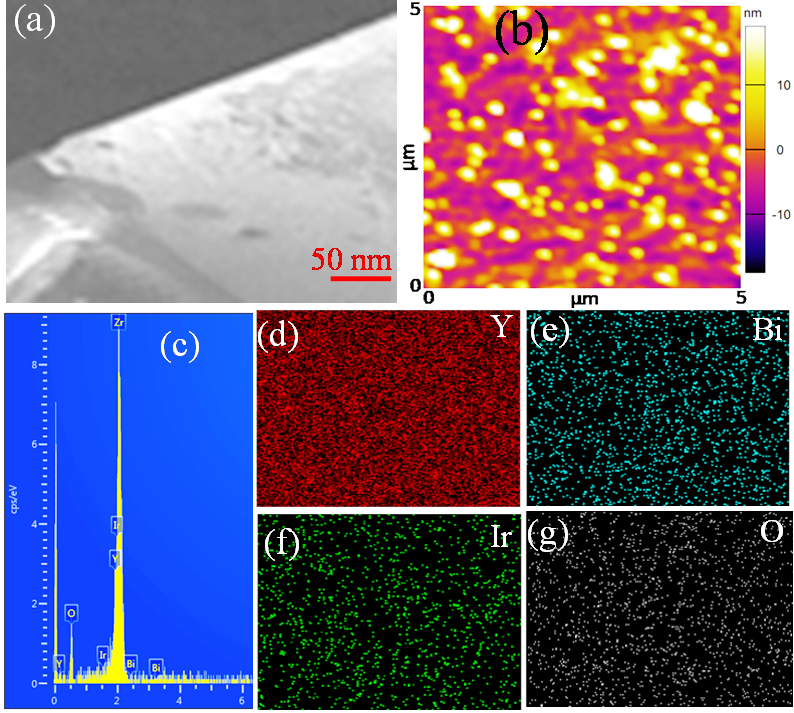}\\
\caption{(a) SEM image around the edge of the film S1 formed due to shadow masking S2. (b) AFM image of the film S1; (c) corresponding EDAX spectra; and elemental mapping for (d) Y, (e) Bi, (f) Ir, (g) O.}\label{fig:semafm}
\end{figure}

SEM micrograph of the thin film S1 is shown in Fig.~\ref{fig:semafm}a. A closer look at the micrograph shows that the surface of the film is crack free and smooth over large area (Fig.~\ref{fig:semafm}a). The average thickness of the film turns out to be $\sim$ 25 nm . The corresponding EDS and elemental mapping suggest satisfactory spatial homogeneity of chemical composition in S1 (Fig.~\ref{fig:semafm}c-g). Figure~\ref{fig:semafm}(b) shows the AFM image over 5 $\mu m^2$ area of film S1. The measured large area RMS roughness of the film is 6.0 nm.

\begin{figure}
\includegraphics[width=9cm]{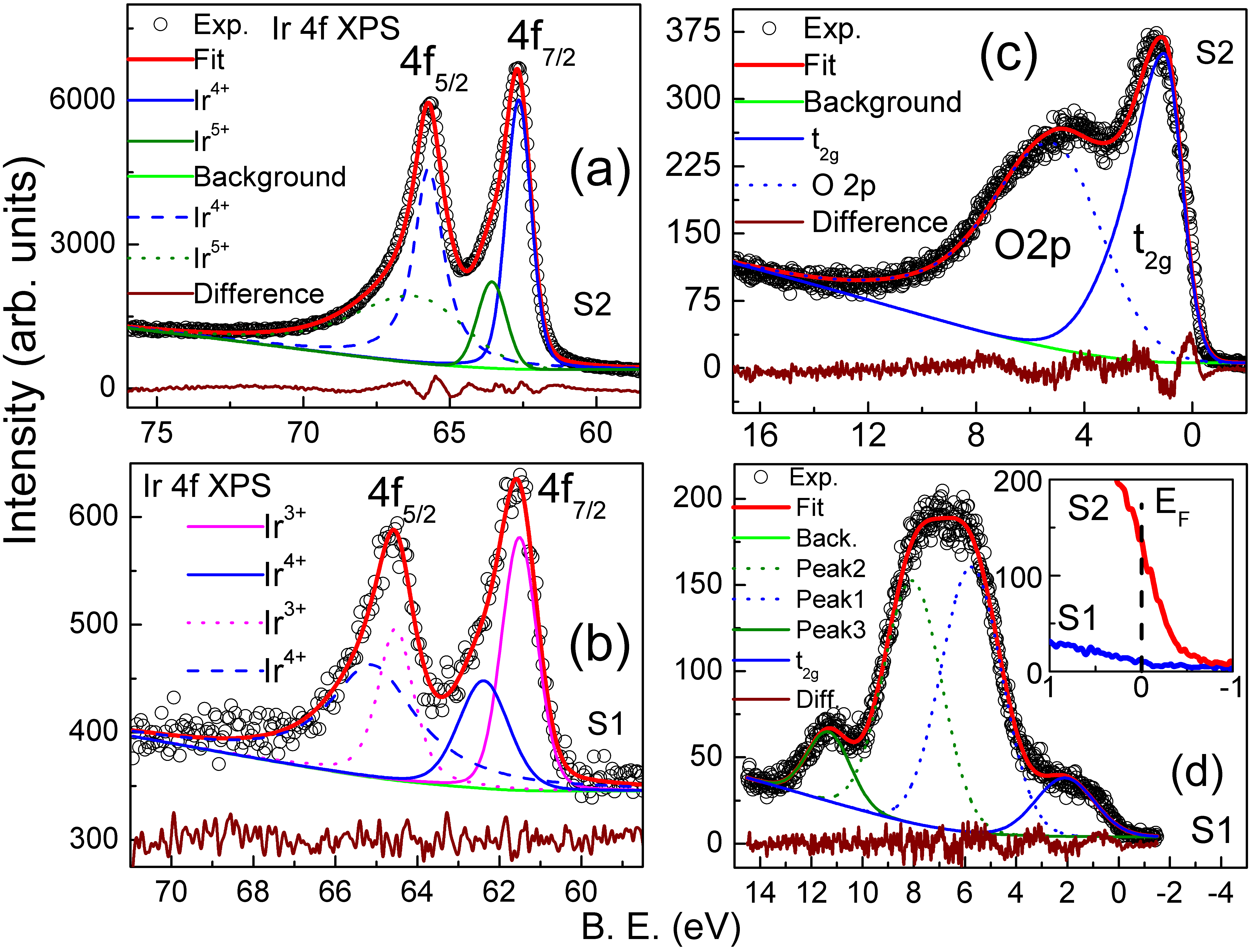}\\
\caption{(a) Ir 4f XPS spectrum of sample S2; (b) The same for S1; (c) Ir 5d valence band XPS spectrum of S2; (d) The same for S1, inset: Density of states for S1 and S2 near the Fermi level.}\label{fig:XPS}
\end{figure}

Fig.~\ref{fig:XPS}a,b displays the XPS spectra for S2 and S1, respectively. We fitted the high-resolution XPS spectra using two components with the help of XPS peakfit4.1 software using the method reported elsewhere~\cite{Yang1}. The observed binding energies corresponding to the two Ir-$4f_{7/2}$ peaks are found to be 61.4 eV and 62.5 eV, which are labeled as $Ir^{3+}$ [solid magenta line in Fig.~\ref{fig:XPS}b] and $Ir^{4+}$ [solid blue line in Fig.~\ref{fig:XPS}a,b], respectively. While the peaks centered at binding energies of 64.6 eV and 65.8 eV are attributed to Ir-$4f_{5/2}$ {$Ir^{3+}$ [dotted magenta line in Fig.~\ref{fig:XPS}b], $Ir^{4+}$ [dashed blue line in Fig.~\ref{fig:XPS}a,b], respectively. On the other hand, the peaks located at binding energy 63.7 eV (solid olive line) and 67 eV (dotted olive line) represent the $Ir^{5+} $ oxidation state of Ir-$4f_{7/2}$ and Ir-$4f_{5/2}$, respectively of sample S2 shown in Fig.~\ref{fig:XPS}a. The XPS spectra for S1 are shifted towards lower binding energies in comparison to S2 ($Ir^{4+}$) XPS spectra (Fig.~\ref{fig:XPS}a,b). The peak area ratio $Ir^{3+}:Ir^{4+}$ calculated from the XPS fitting is 0.77:1, suggesting near equal distribution of $Ir^{3+}$ and $Ir^{4+}$ sites. The origin of mixed oxidation states is likely due to oxygen deficiency. Existence of lower oxidation states in Ir based pyrochlore compounds have been reported before~\cite{Yang1,Yang2,Jafer}. Similarly, the Y 3d and Bi 4f core level XPS spectra (not shown here) shows superposition of the peaks due to overlapping~\cite{Jafer} of binding energy ranges suggesting only $3+$ oxidation states for Bi and Y cations.

We measured the XPS valence-band spectra (VBS) of S1 and S2 as shown in Fig.~\ref{fig:XPS}d,c. The peaks located at the binding energy 1.5 eV and 6 eV are assigned as $t_{2g}$ and $O(2p)$, respectively. We observe suppression in density of states associated with $t_{2g}$ orbital in film S1 compared to polycrystalline sample S2 (shown in the inset), which is consistent with the higher resistivity and semiconducting nature of the film. The suppression of $t_{2g}$ peak close to $E_F$ region in S1 might be due to the contribution mostly from the completely filled $J_{eff}=3/2$ and $J_{eff}=1/2$ level. The electronic structure of $Ir^{4+}$ and $Ir^{3+}$ are $5d^5$ and $5d^6$, respectively. The resulting fully filled $J_{eff} = 1/2$ level corresponding to $Ir^{3+}$ makes the system non-magnetic and less conductive in contrast to the case reported earlier for nanocrystalline iridates~\cite{Vinod} where the partially filled $J_{eff}=1/2$ levels due to coexistence of $Ir^{4+}$ and $Ir^{5+}$ makes the system highly conducting and magnetic. Overall, VBS analysis shows suppression in $t_{2g}$ density of states close to the Fermi level in thin film which is otherwise very prominent in the polycrystalline sample (inset, Fig.~\ref{fig:XPS}d) .

\begin{figure}
\includegraphics[width=9cm]{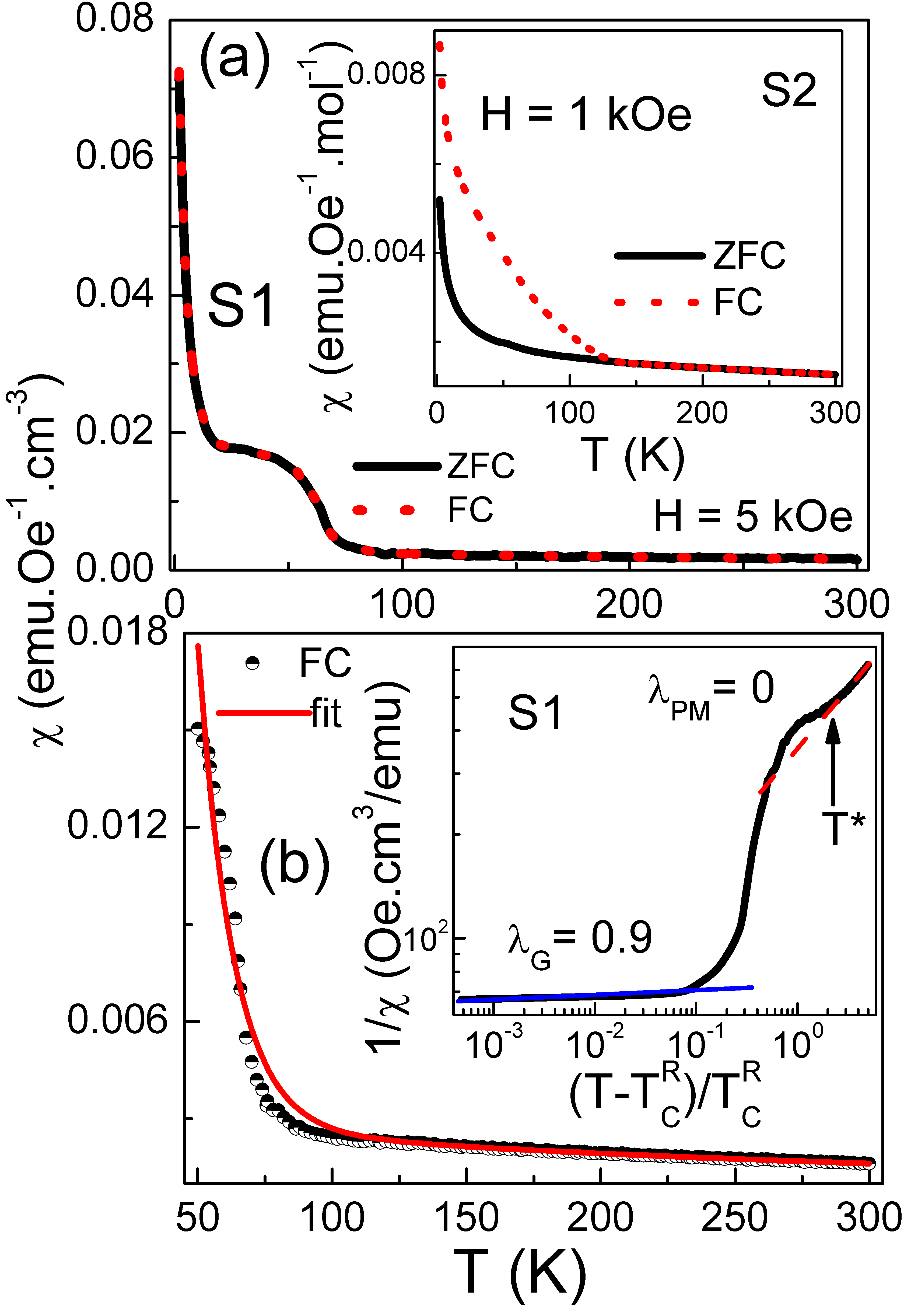}\\
\caption{(a) In-plane magnetic susceptibility as a function of temperature for S1; and S2 (inset). (b) The solid red line demonstrate the theoretical curve obtained within the order of the Griffiths-phase model; inset shows log-log plots of $\chi^{-1}$ versus $\frac{T-T^R_C}{T^R_C}$ (reduced temperature) for S1.}\label{fig:MT}
\end{figure}

\begin{figure}
\includegraphics[width=9.0cm]{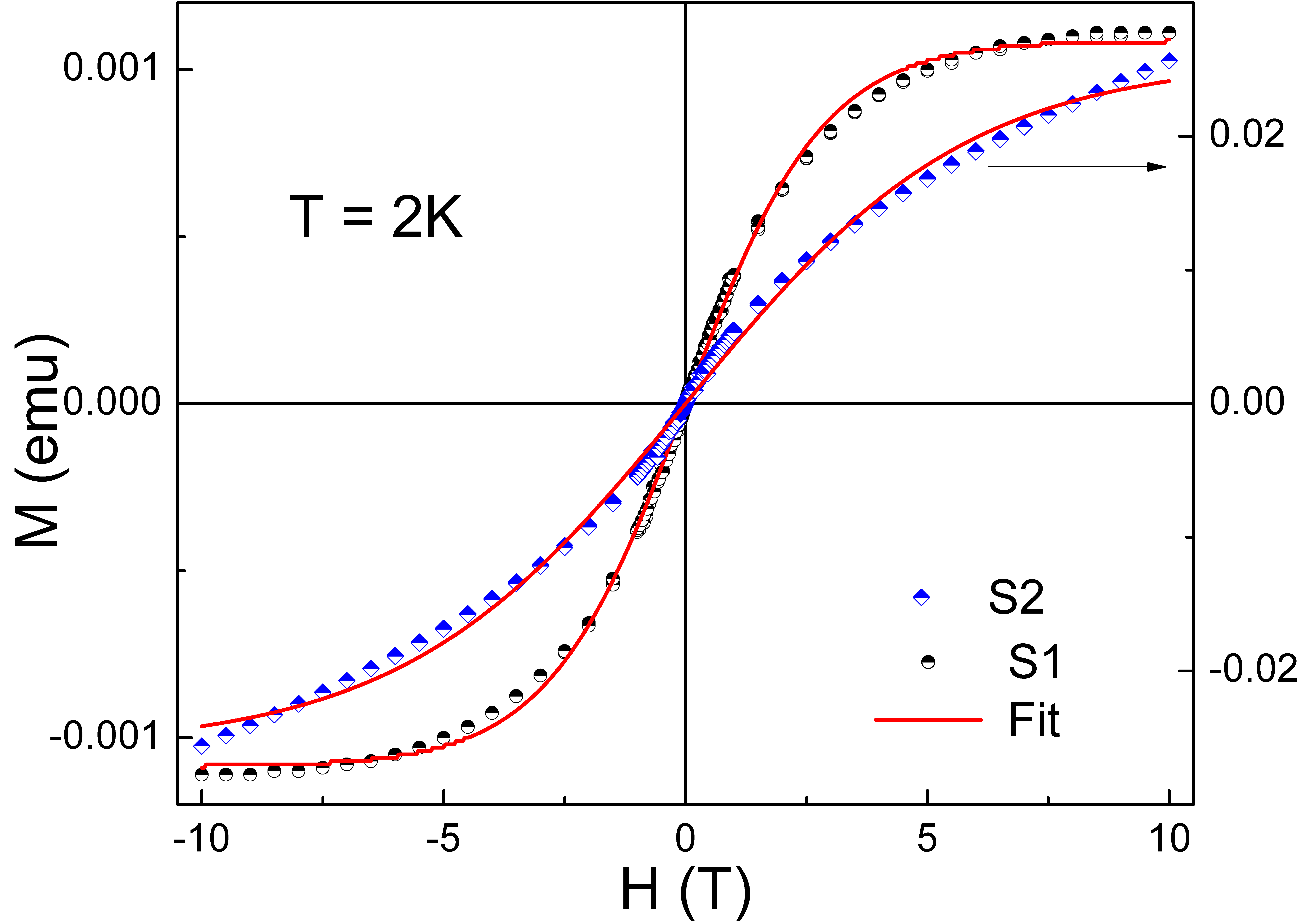}\\
\caption{Magnetization as a function of applied magnetic field for S1 and S2. The solid red line is the fit of 2K M(H) data based on the Brillouin function according to Eq.~\ref{Eq.Brill}.}\label{fig:MH}
\end{figure}

\begin{figure}
\includegraphics[width=9cm]{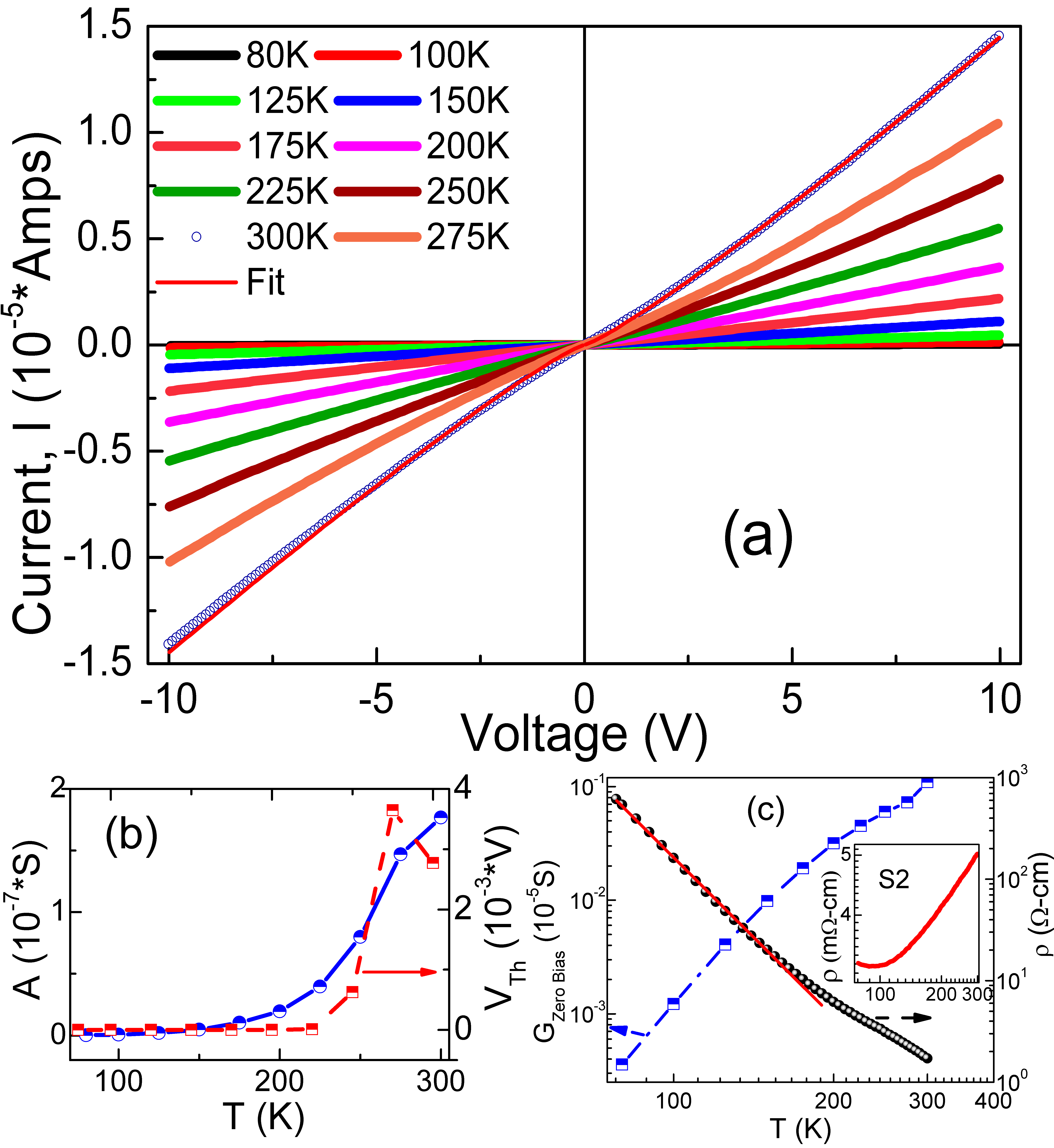}\\
\caption{(a) I vs V isotherms at selected temperatures with LZB model fit shown for 300K. Inset: Fit parameters $A$ (blue line) and $V_{Th}$ (red line) as a function of temperatures; b) Zero bias conductance and resistivity as function of temperature; inset shows resistivity versus temperature of S2.}\label{fig:IV}
\end{figure}

\begin{figure}
\includegraphics[width=8cm]{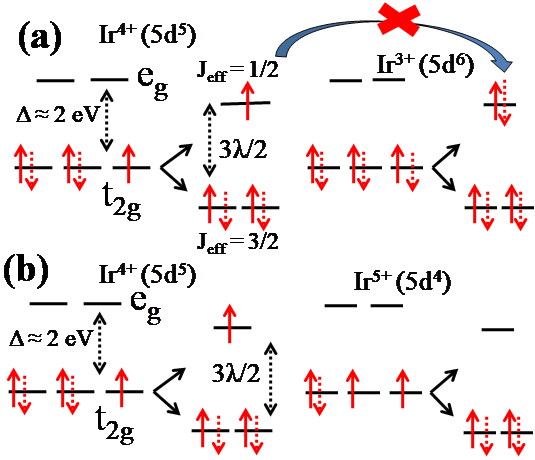}\\
\caption{(a) A schematic demonstration of spin selective hopping of $J_{eff}=1/2$ level electron between $Ir^{3+}$ and $Ir^{4+}$ combination. (b) In this combination electron can hop from $J_{eff}=1/2$ level electron of $Ir^{4+}$ to $Ir^{5+}$.}\label{fig:GP}
\end{figure}

Figure~\ref{fig:MT}(a) shows the zero field cooled (ZFC) and field cooled (FC) temperature dependent magnetic susceptibility ($\chi = M/H$) of films (S1) at applied in-plane magnetic field H = 5 kOe after eliminating the diamagnetic contribution due to the substrate. The change in slope at high temperature and sharp upturn at low temperature in $\chi(T)$ curve is qualitatively similar to that observed in S2 and several other pyrochlore iridate systems~\cite{Yanagishima,Aito,Soda,Liu,Taira,Harish,Zhu,Vinod}. However, strikingly, the bifurcation between ${\chi_{ZFC}}$ and ${\chi_{FC}}$ branches commonly observed for S2 [inset of Fig.~\ref{fig:MT}(a)] is completely absent in S1, suggesting suppression of long range magnetic ordering, possibly due to enhanced magnetic frustration. The magnetic susceptibility in the paramagnetic (PM) regime can be well described using the Curie-Weiss law, $\chi = \frac{C_{CW}}{T-\theta_{CW}}$, where $C_{CW}$ is the Curie constant, $\theta_{CW}$ is the Curie-Weiss temperature. The obtained fitting parameters using Curie-Weiss law is given in Table~\ref{tab:Brill}. The negative value of $\theta_{CW} \sim$ -315K for thin film S1 is almost half compared to the bulk sample S2 ($\theta_{CW}\sim$ -615K), suggesting reduction of anti-ferromagnetic (AFM) correlation. The calculated effective magnetic moments in paramagnetic phase~\cite{Blundell,Feng} turn out to be $\mu_{eff}^{M(T)}$ = 1.5$\mu_B$ and 0.9$\mu_B$ for S1 and S2, respectively. The value of $\mu_{eff}^{M(T)}$ for S1 is very close to the predicted theoretical value [$\mu^{theor}_{eff}$ = 1.73$\mu_B$]. The in-plane magnetization as a function of applied magnetic field measured at temperature 2K is shown in the Fig.~\ref{fig:MH}. M(H) curve for S1 shows hint of saturation at higher magnetic field suggest field induced ordering of magnetic moments which is absent in S2 up to applied magnetic field 10T. The observed hysteresis is negligible. We analyzed the M(H) data using the Brillouin function as well~\cite{Blundell}. For a pyrochlore iridate the low temperature magnetic state can be analysed with an effective spin J = 1/2 where

\begin{equation}\label{Eq.Brill}
\frac{M}{M_S} = B_{1/2}(y)
\end{equation}
where $y = \left(\frac{g_J \mu_B J}{k_BT}\right)H$, $g_J$ is the Lande' g-factor, and $\mu_B$ is Bohr magneton. The value of effective magnetic moment $\mu_{eff}^{M(H)}$ is estimated using $\mu_{eff} = g_J\mu_B\sqrt{J(J+1)}$, where $g_J$ is extracted from the Brillouin function fit. The obtained fitting parameters of M(H) data are given in Table~\ref{tab:Brill}. While the calculated effective magnetic moment from M(H) data i.e. $\mu_{eff}^{M(H)}$ for film and its bulk counterpart is very close to the effective magnetic moment $\mu_{eff}^{M(T)}$ determined from the temperature dependent magnetization data, the $\mu_{eff}^{M(H)}$ for film is greater than the expected theoretical value [$\mu^{theor}_{eff}$ = 1.73$\mu_B$]. For S2 the experimental data deviates considerably from the Brillouin function fit over the entire field range [Fig.~\ref{fig:MH}], suggests non-paramagnetic behaviour.  

\begin{table}[h!]
\centering
\caption{Parameters obtained from fitting of the $M(T)$ and M(H) data using Curie-Weiss law and Brillouin function, respectively.}\label{tab:Brill}
\begin{tabular}{|c|c|c|c|c|}
 \hline
Sample &$\theta_{CW}$& $\mu_{eff}^{M(T)}$& $g_J$ & $\mu_{eff}^{M(H)}$\\
 &(K)&($\mu_B$)& &($\mu_B$)\\
 \hline
 S1(Film) & -315 & 1.5 & 2.20 & 1.90 \\
 \hline
 S2(Poly) & -615 & 0.9 & 1.06 & 0.92 \\
 \hline
 \end{tabular}
\end{table}

Although we do not observe any signature of long range ordering in S1, the sharp downward deviation from the Curie-Weiss behavior [inset of Fig.~\ref{fig:MT}(b)] in the temperature dependence of FC dc inverse susceptibility strongly suggests formation of magnetically ordered rare regions below $T^\ast \sim 168K$ quite in contrast to the bulk polycrystalline sample. Such type of magnetic behaviour is typical of percolative Griffiths-like phase~\cite{Griffiths}. Within the Griffiths phase scenario, the magnetic ordering temperature, $T_C(x)$, of a randomly diluted ferromagnet is lower than the same of the undiluted one ($T_C^{x=0}$). The magnetization will be a non-analytic function in this region i.e. $T_C(x) \leq T \leq T_C^{x=0}$ due to the formation of low density clusters with short-range correlations. The Griffiths temperature $T^\ast$ marks the point of deviation from ideal paramagnetic CW behaviour. The magnetic phase between the temperature range $T_{C}$ and $T^\ast$ corresponds to the Griffiths phase regime. In this regime the magnetic phase is neither a PM nor an ideal long range FM, instead the magnetic clusters are embedded inside the PM matrix. We analyze the presence of short range FM correlations in PM state by fitting the temperature dependence of inverse dc magnetic susceptibility following the protocol discussed elsewhere~\cite{Vinay,Ouyang,Rathi,Sanjib1,Phong,Sanjib2} and using the relation $\chi^{-1} \propto (T - T_C^R)^{1-\lambda}$, where $0 \leq \lambda \leq 1$ and $T_C^R$ is the critical temperature of random ferromagnetic clusters. The value of $T_C^R$, obtained such that $\lambda=0$ above $T^\ast$, turns out to be 50 K. Using the logarithmic plot shown in inset of Fig.~\ref{fig:MT}(b), the power law exponent of the value $\lambda=0.9$ is extracted from the slope of the low temperature side of the experimental data below $T^\ast$. The effective magnetic moment of FM cluster can be estimated by fitting the temperature dependence of the magnetic susceptibility above 50K shown in Fig.~\ref{fig:MT}(b) with following expression $\chi(T,H) = \chi_0^G exp(-k_BT/\mu_{eff}H) + C_{CW}/(T+\theta_{CW})$, where $\chi_0^G$ is susceptibility of the GP at T = 0K and $\mu_{eff}$ is the effective magnetic moment of the ferromagnetic cluster~\cite{Prokhorov,Hyun,Solin}. The estimated value of effective magnetic moment of FM cluster within the framework of this model $\mu_{eff}^{FM}$ = 38.5$\mu_B$. The estimation of effective magnetic moment $\mu_{eff}$ gives information about the formation of Griffiths-like ferromagnetic (FM) clusters. On the other hand, cluster size or correlation length can be estimated using small angle neutron scattering~\cite{Magen} spectra and the fraction of spins contributing to cluster formation can be estimated by electron paramagnetic resonance spectroscopy~\cite{Vinay} which are beyond the scope of this article. The calculated value of $\mu_{eff}^{M(T)}$ above $T^\ast$ from the slope of $\chi^{-1}(T)$ curve is close to the theoretical value of spin only moment $\mu_{eff}^{theor}$. Strikingly, we observe huge enhancement in $\mu_{eff}^{FM} = 38.5 \mu_B$ below $T^\ast$ much greater than the theoretical value. This suggests the formation of presumably nanosized Griffiths-like ferromagnetic (FM) clusters.

The isothermal I-V characteristic curves for the film measured within the temperature range 80-300 K are shown in Fig.~\ref{fig:IV}a. Since the valence band spectra of the thin film S1 clearly shows opening of a gap at the fermi level, the observed non-linear I-V curves can be analyzed within the Landau-Zener breakdown (LZB) model~\cite{Hardy} related to electric field induced mobile excitation of the charge carrier overcoming the correlation gap. The current is given by, $I = -AV\ln(1-e^{{-V_{Th}}/{V}})$, where, $V_{Th}$ and $A$ represent threshold voltage (above which the current increases non-linearly) and scaling factor which sets the overall curve slope, respectively. A representative fit to the experimental I(V) curve is shown for 300K in Fig.~\ref{fig:IV}a. The obtained best fitting parameters $A$ and $V_{Th}$ are plotted against temperature as shown in Fig.~\ref{fig:IV}b. The zero bias conductance ($G_{0}$) for S1 increases as temperature increases suggesting semiconducting behavior (Fig.~\ref{fig:IV}c). The temperature dependence of resistivity ($\rho$) for S1 fitted with power law model (solid red line) with exponent $n=3.8$ is shown in Fig.~\ref{fig:IV}c. This is in complete contrast to the bulk sample S2, which shows metallic behaviour with a shallow low temperature minimum (inset, Fig.~\ref{fig:IV}c).

The suppression of electrical conduction in the thin film S1 might be attributed to the reduction in $t_{2g}$ DOS along with coexistence of non-magnetic $Ir^{3+}$ and $Ir^{4+}$ ions as revealed by the XPS data. The Bi 6s/p orbital lies near the Fermi level and hybridizes strongly with the 5d orbital of Ir, making the Ir 5d bandwidth wider thus leading to the enhanced conductivity in S2. On the other hand, in S1, the hopping probability decreases due to the $Ir^{4+}-O^{2-}-Ir^{3+}$ configuration. The $Ir^{3+}$ site corresponds to the filled $J_{eff}=1/2$ level (Fig.~\ref{fig:GP}). Neglecting higher order processes, a down spin corresponding to the $J_{eff}=1/2$ level at the $Ir^{3+}$ site is allowed to hop to the nearest neighbour $Ir^{4+}$ site while the up spin cannot hop because the nearest neighbour up spin state is already occupied. On the other hand, hopping from $Ir^{4+}$ site to $Ir^{3+}$ is blocked regardless of the spin state. Virtual hopping between the two occupied up spin states is, however, allowed, leading to enhanced FM correlation. Nonetheless, presence of lower oxidation states could be one of the major reasons behind the suppressed  electrical conduction.

The question one now needs to answer is the following: how does the presence of non-magnetic defects in the network of Ir tetrahedra give rise to FM clusters below $T^\ast$? Above $T^\ast \sim 168 K$ i.e. $T \geq T^\ast$ the system is in paramagnetic phase. When the system is cooled down, FM correlated clusters appear for $T \leq T^\ast$. In an ideal pyrochlore iridate, magnetic $Ir^{4+}$ ions form a corner-sharing tetrahedral network where the localized spins point either into or outward from its center forming an all-in/all-out AFM ordered phase. In the presence of non-magnetic defects in the form of $Ir^{3+}$ sites, the six $Ir^{4+}$ moments surrounding a defect can be visualized as a FM or canted AFM cluster. In such a scenario, virtual hopping of similar spins between two nearest neighbour sites with different oxidation states should lead to enhanced FM correlation. Such cation antisite defects which lead to magnetic disorder in the system could be due to either Ir deficiency or oxygen vacancies. The resulting magnetic disorder divides the undiluted system into small rare regions. It should be noted that the FM correlations in the thin film is unusually strong. The temperature scale $T^\ast$ ($\sim$ 168K) is much greater than the $T^R_C$ ($\sim$ 50 K) and consequently, the range of the GP given as $T^\ast - T^R_C/T^R_C = 2.36$, is much larger than that for other systems which exhibit Griffiths like phase~\cite{Ouyang}.

\section{CONCLUSIONS}

To summarize, we find that the transport and magnetic properties of $Y_{1.7}Bi_{0.3}Ir_2O_7$/YSZ(100) thin film prepared by pulsed laser deposition is qualitatively different from its bulk counterpart. Instead of long range magnetic ordering as is characteristic of the bulk, we observe features similar to the ferromagnetic Griffiths phase in thin film spanning a large temperature range. The XPS study shows coexistence of mixed oxidation states of Ir, [i.e. $Ir^{4+}$ and $Ir^{3+}$], with the lower oxidation state suggesting oxygen deficiency or non-magnetic defects. The valence band spectra show significant reduction in density of states at Fermi level in the thin film compared to the polycrystalline sample, which is consistent with the observed suppressed electrical conductivity in the former. We further emphasize that existence of lower oxidation state of Ir in thin film might be responsible for the formation of FM rare regions as well as suppressed electrical conduction.

\end{document}